\documentclass[pra,twocolumn,aps,superscriptaddress,showpacs,floatfix]{revtex4}

\usepackage{amsmath}
\usepackage{amssymb}
\usepackage{amsbsy}
\usepackage{graphicx}
\usepackage{color}
\usepackage{epstopdf}
\usepackage{mathrsfs}

\begin{document}

\title{Fate of zero modes in a finite Su-Schrieffer-Heeger model with $\mathcal{PT}$ symmetry}

\author{Zhihao Xu}
\affiliation{Institute of Theoretical Physics, Shanxi University, Taiyuan 030006, China}
\affiliation{Collaborative Innovation Center of Extreme Optics, Shanxi University, Taiyuan 030006, P.R.China}
\affiliation{State Key Laboratory of Quantum Optics and Quantum Optics Devices, Institute of Opto-Electronics, Shanxi University, Taiyuan 030006, P.R.China}
\email{xuzhihao@sxu.edu.cn}
\author{Rong Zhang}
\affiliation{Institute of Theoretical Physics, Shanxi University, Taiyuan 030006, China}
\affiliation{Collaborative Innovation Center of Extreme Optics, Shanxi University, Taiyuan 030006, P.R.China}
\author{Shu Chen}
\affiliation{Beijing National Laboratory for Condensed Matter Physics,
Institute of Physics, Chinese Academy of Sciences, Beijing 100190, China}
\affiliation{School of Physical Sciences, University of Chinese Academy of Sciences, Beijing, 100049, China}
\affiliation{Yangtze River Delta Physics Research Center, Liyang, Jiangsu 213300, China}
\author{Libin Fu}
\affiliation{Graduate School of China Academy of Engineering Physics, No. 10 Xibeiwang East Road, Haidian District, Beijing, 100193, China}
\author{Yunbo Zhang}
\affiliation{Institute of Theoretical Physics, Shanxi University, Taiyuan 030006, China}
\affiliation{Key Laboratory of Optical Field Manipulation of Zhejiang Province and Physics Department of Zhejiang Sci-Tech University,	Hangzhou 310018, China}

\begin{abstract}
Due to the boundary coupling in a finite system, the zero modes of a standard Su-Schrieffer-Heeger (SSH) model may deviate from exact-zero energy. A recent experiment has shown that by increasing the system size or altering gain or loss strength of the SSH model with parity-time ($\mathcal{PT}$) symmetry, the real parts of the energies of the edge modes can be recovered to exact-zero value [Song \emph{et al.} Phys. Rev. Lett. \textbf{123}, 165701 (2019)]. To clarify the effects of $\mathcal{PT}$-symmetric potentials on the recovery of the nontrivial zero modes, we study the SSH model with $\mathcal{PT}$-symmetric potentials of different forms in both infinite and finite systems. Our results indicate that the energies of the edge modes in the infinite size case decide whether or not the success of the recovery of the zero modes by tuning the strength of $\mathcal{PT}$-symmetric potential in a finite system. If the energies of the edge modes amount to zero in the thermodynamic limit under an open boundary condition (OBC), the recovery of the zero modes will break down by increasing the gain or loss strength for a finite system. Our results can be easily examined in different experimental platforms and inspire more insightful understanding on nontrivial edge modes in topologically non-Hermitian systems.
\end{abstract}


\maketitle
\section{Introduction}

Recently, non-Hermitian systems have been greatly studied both in experimental and theoretical fields \cite{Bender,Bender1,Hatano1,Hatano2,Hatano3,QBZeng1,SLongi1,SLongi2,SLongi3,HuiJiang,QBZeng2,HuitaoShen,Shunyu1,Shunyu2,Zongping,XZZhang,Lee1,Lee2,Harari,Parto,HengyunZhou,Ruter,LiangFeng,Regensburger,Joglekar,XiangZhan,LeiXiao,Zeuner,Hirsbrunner,YongXu,YuChen,Okuma,LinhuLi,DengTianShu,Brzezicki,Carlstrom,Budich,Kunst,Takata,MPan,Nakagawa,Hamazaki,Yamamoto,Ashida,Kawabata,Tomita,Kawabata1,Kawabata2,LihongZhou,ChuanhaoYin,Yoshida1,Yoshida2,Yoshida3,Yoshida4,JinLiang3}. Specially, traditional topological phases considered in closed systems described by Hermitian Hamiltonians have permeated into open systems governed by non-Hermitian operators. The non-Hermitian descriptions arise when the system interacts with an environment \cite{Daley,Dalibard}. It has been revealed that non-Hermiticity can greatly alter the topological behaviors that were established in the Hermitian cases, such as the failure of the bulk-boundary correspondence \cite{HuitaoShen,Shunyu1,Shunyu2,Zongping,Lee1,Takata,Kawabata1,JinLiang2}, the skin effect \cite{SLongi1,HuiJiang,Shunyu1,Shunyu2,Zongping}, the boundary-dependent spectra \cite{Hatano2,Kunst, Kawabata1,Longhi4,Yuce2}, the non-Hermitian-induced topology \cite{Takata,MPan}, and the edge modes influenced by gain and loss distributions \cite{BaogangZhu,Klett,Yuce,Yuce1,JinLiang,KentaEsaki,DanielLeykam,WangeSong,JinLiang1}.

On the other hand, researching on topological states of matter traced back to over three decades ago has attracted considerable interest in many fields of physics, including photonics \cite{Rechtsman,Plotnik,Joannopoulos,Maczewsky,Mukherjee}, cold atomic gases \cite{Jotzu, Aidelsburger}, acoustic systems \cite{ZhaojuYang,Khanikaev}, and mechanics \cite{Susstrunk,Nash}. A topological insulator exhibits an insulating bulk and gapless edge states under  open boundary conditions (OBCs), which can be characterized by a topological invariant \cite{Bernevig,Hasan,XiaoliangQi}. A series of landmark topological models have been experimentally realized, such as the Su-Schrieffer-Heeger (SSH) model \cite{MPan,Atala1,WangeSong,St-Jean,Meier}, the Haldane model \cite{Rechtsman,Jotzu}, and the Hofstadter model \cite{Aidelsburger}. Particularly, the one-dimensional SSH model may be the simplest two band topological system initially introduced to study the polyacetylene \cite{SSH}. The chiral symmetry of the SSH model leads to nontrivial topology which can be probed by the winding number and the presence or absence of two-fold degenerate zero modes in the thermodynamic limit under OBC. In fact, the experimental realization of the topological systems is usually of finite size. Due to the boundary coupling in a finite system, the zero modes will deviate from exact-zero energy \cite{Ryu}. Song \emph{et al.} \cite{WangeSong} suggested that one can reduce the coupling of the boundary modes and recover the zero modes in a finite system by increasing the  alternating gain or loss strength in a one-dimensional parity-time ($\mathcal{PT}$)-symmetric SSH model.

An interesting issue that arises here is whether the topological zero modes can always be recovered by the $\mathcal{PT}$-symmetric potential in a small system case. To answer the question, we consider the SSH model with different types of $\mathcal{PT}$-symmetric potentials and study the breakup and recovery of the nontrivial zero modes by controlling the gain or loss strength. We focus on the effect of $\mathcal{PT}$-symmetric potentials on the edge modes in both large and small size limits. Our results indicate that if the energies of the nontrivial edge modes of a $\mathcal{PT}$-symmetric SSH model amount to zero in a large system, the modes shall not be recovered by tuning the gain or loss strength in a finite system.

The paper is organized as follows. In Sec. II, we present the Hamiltonian of a SSH model with $\mathcal{PT}$-symmetric potentials of different forms. In Sec. III, we study the edge modes of the SSH model with an alternating gain or loss potential and a pair of balanced gain and loss at two end sites in both large and small size cases. In both cases, we find the zero modes can be recovered by modulating the gain or loss strength. In Sec. IV, we discuss the failure of the recovery of the edge modes for the SSH model with another types of $\mathcal{PT}$-symmetric potentials in the small size limit. To further clarify our results, we discuss the edge modes of a spin-orbit coupled SSH chain with $\mathcal{PT}$-symmetry in Sec. V. Finally, a conclusion is given in Sec. VI.

\section{SSH model with $\mathcal{PT}$ symmetry}

\begin{figure}[tbp]
	\begin{center}
		\includegraphics[width=.45 \textwidth] {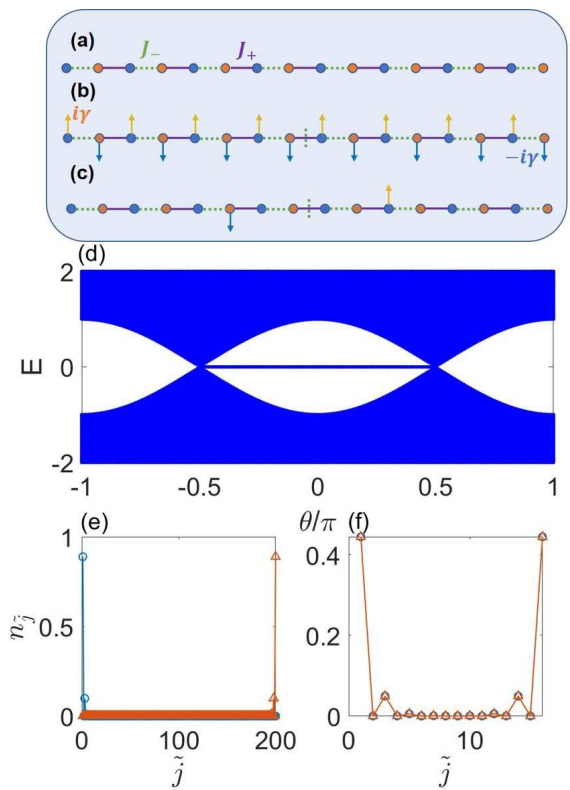}
	\end{center}
	\caption{(Color online) (a) Schematic diagram of the conventional SSH model. (b,c) Sketch of the SSH model with the gain and loss potentials of different forms. Here, $J_{-}$ is denoted by the green dashed line, $J_{+}$ is denoted by the purple solid line, the up (down) arrow represents the gain (loss) at a lattice site with amplitude $\gamma$. (d) Energy spectrum of SSH model changing with $\theta$ for $J=1$, $\Delta=0.5$ and $N=100$ under OBC. (e) Density distributions of two exact-zero modes for $J=1$, $\theta=0, \Delta=0.5$ and $N=100$. (f) Density distributions of two near-zero modes for $J=1$, $\theta=0, \Delta=0.5$ and $N=8$. Here, the horizontal coordinate is the position of the lattice site $\tilde{j}=2(j-1)+1$ for one on $A$ sublattice in the $j$-th unit cell and $\tilde{j}=2j$ for one on $B$ sublattice in the $j$-th unit cell.}\label{Fig1}
\end{figure}

We consider a one-dimensional non-Hermitian SSH model with alternating modulated hopping parameters whose imaginary part is distributed in a manner that presents $\mathcal{PT}$ symmetry, that is, the loss and gain potentials of different forms. The Hamiltonian can be described as
\begin{equation}\label{eq1}
\hat{H}^{(\beta)}=\hat{H}_0+\hat{H}_1^{(\beta)},
\end{equation}
with
\begin{equation}\label{eq2}
\hat{H}_0=\sum_{j}(J_{-}\hat{c}_{j,A}^{\dagger}\hat{c}_{j,B}+J_{+}\hat{c}_{j,B}^{\dagger}\hat{c}_{j+1,A}+H.c.),
\end{equation}
and the $\mathcal{PT}$-symmetric potentials $\hat{H}_1^{(\beta)}$.
For the conventional SSH model, the Hamiltonian can be written as $\hat{H}^{(\beta)}$ with $\beta=a$, and $\hat{H}_1^{(\beta)}=0$ and there are two sublattices in each unit cell marked by $A$ and $B$ which is shown in Fig. \ref{Fig1}(a). $\hat{c}_{j,\alpha}$ is the annihilation operator for particles on the $\alpha$-th lattice in the $j$-th unit cell and $J_{\pm}=J(1\pm \Delta \cos{\theta})$ are the alternating hopping strengths with $J=1$ being set as the unit of energy, the dimerization strength is $\Delta$, and $\theta$ denotes the tunneling parameter. As shown in Fig. \ref{Fig1}(a), the amplitude of intracell tunneling $J_{-}$ is denoted by the green dashed line and the purple solid line denotes intercell tunneling energy $J_{+}$. For convenience, we consider $\Delta\in [0,1)$ and $\theta$ can vary from $-\pi$ to $\pi$. The non-Hermitian term $\hat{H}_1^{(\beta)}$ with $\beta=\{b,c\}$ represents the gain and loss potentials of different forms corresponding to those shown in Figs. \ref{Fig1}(b) and \ref{Fig1}(c), respectively, to keep the total Hamiltonian $\hat{H}$ carrying an additional $\mathcal{PT}$ symmetry, $\mathcal{PT}\hat{H}(\mathcal{PT})^{-1}=\hat{H}$. Here, the parity operator $\mathcal{P}$ satisfies $\mathcal{P}\{j,\alpha\}\mathcal{P}^{-1}=\{N+1-j,\bar{\alpha}\}$, where $N$ is the number of the unit cell and if $\alpha=A (B)$, $\bar{\alpha}=B (A)$; and $\mathcal{T}$ is the time-reversal operator which satisfies $\mathcal{T} i\mathcal{T}^{-1}=-i$. The $\mathcal{PT}$-symmetric potentials of different forms are described as following: (b) $\hat{H}_1^{(b)}=\sum_{j,\alpha}(-1)^{\eta}i\gamma\hat{c}_{j,\alpha}^{\dagger}\hat{c}_{j,\alpha}$ and (c) $\hat{H}_1^{(c)}=\sum_{j,\alpha}i\gamma[(-1)^{\eta}\delta_{j,n}\delta_{\alpha,\alpha^{\prime}}+(-1)^{\eta+1}\delta_{j,N-n+1}\delta_{\alpha,\bar{\alpha}^{\prime}}]\hat{c}_{j,\alpha}^{\dagger}\hat{c}_{j,\alpha}$ with $\gamma$ being the gain or loss amplitude, $n\le N/2$ and $\eta=0(1)$ for $A$($B$) sublattice. Here, case (b)  corresponds to an alternating gain or loss effect of the whole chain and case (c) corresponds to two conjugated imaginary defect potentials to keep the SSH model with $\mathcal{PT}$ symmetry. In Fig. \ref{Fig1}, the gain (loss) strength on the $\alpha$-th lattice in the $j$-th unit cell which is represented by up (down) arrow can be described as $i\gamma$ ($-i\gamma$).

In the absence of the non-Hermitian term [case (a), $\hat{H}_1^{(a)}=0$], the conventional SSH model as the simplest two-band topological system describes a chiral chain of the BDI symmetry class which obeys $\Gamma \hat{H}^{(a)} =-\hat{H}^{(a)}\Gamma$. Here, the chiral operator $\Gamma=\bigoplus_{j=1}^{N}\sigma_z$ represents the $z$-component Pauli operator $\sigma_z$ acting on the internal Hilbert space of each unit cell. The chiral symmetry in a one-dimensional system supports a topologically nontrivial phase with the nontrivial winding number $W=1$ and doubly degenerate edge modes for large-enough size under OBC. The topologically nontrivial regime is $\theta\in(-\pi/2,\pi/2)$ for the SSH chain and the nontrivial zero-mode edge states are found in such phases, which are quite different from the bulk one with extended distributions. However, there are no edge states in the regime $\theta\in [-\pi,-\pi/2)$ and $(\pi/2,\pi]$ corresponding to the trivial winding number $W=0$ seen in Fig.\ref{Fig1}(d) with $\Delta=0.5$ and $N=100$. To study the edge modes under OBC, we consider a single-particle state, $\psi(E)=\sum_{j,\alpha}\varphi_{j,\alpha}(E)\hat{c}_{j,\alpha}^{\dagger}|0\rangle$ with an elementary excitation energy $E$ and the Schr\"{o}dinger equation can be solved by using the transfer matrix method \cite{Bernevig}. 
In the thermodynamic limit, the energies of the edge modes $E=0$. For $\varphi_{1,B}(0)=0$, the wave function of sublattice $A$ in the $j$-th unit cell is $\varphi_{j,A}(0)=(-\xi)^{j-1}\varphi_{1,A}(0)$ and for $\varphi_{1,A}(0)=0$, $\varphi_{j,B}(0)=(-\xi)^{N-j}\varphi_{N,B}(0)$, where
$\varphi_{1,A}(0)$/$\varphi_{N,B}(0)$ is the initial probability amplitude of the wave function at the left or right edge site with $E=0$ and $\xi=J_{-}/J_{+}$. When $J_{-}=J_{+}$, the zero modes are extended. For the $J_{-}\ne J_{+}$ case, we must restrict $\xi<1$, the wave function of  sublattice $A$ ($B$) shall be localized near the left (right) edge region. Specially, there is no distribution at sublattice $B$ ($A$) for the left (right) edge mode. Figure \ref{Fig1}(e) shows the density distributions $n_{\tilde{j}}$ of the edge modes for $\Delta=0.5$, $\theta=0$, and $N=100$ where $\tilde{j}=2(j-1)+1$ for sublattice $A$ of the $j$-th unit cell and $\tilde{j}=2j$ for sublattice $B$ of the $j$-th unit cell. Both modes are localized at the single-boundary sites. Due to the boundary coupling in a small size system, the degenerate zero modes deviate from the exact-zero energy and their densities have a localized distribution at both boundaries which are shown in Fig. \ref{Fig1}(f) with $N=8$.

For the SSH model with $\mathcal{PT}$ symmetry in our cases, it possesses pseudo-anti-Hermiticity $\Gamma \hat{H}^{(\beta)\dagger} = -\hat{H}^{(\beta)} \Gamma$ with $\beta=\{b,c\}$ which leads to nontrivial topology for non-Hermitian cases \cite{Takata,MPan,SDLiang,Lieu,Schomerus1,Schomerus2,Schomerus3,Schomerus4} and a pair of conjugate imaginary-energy edge states when the edge modes break the $\mathcal{PT}$ symmetry. In the following, we consider the spectra and size-dependent edge modes of SSH model subjected to different gain and loss on-site potentials shown in Figs.\ref{Fig1}(b) and \ref{Fig1}(c).

\section{Recovery of zero modes}

First, we consider the SSH model with an alternating gain or loss shown in Fig. \ref{Fig1}(b) and its Hamiltonian can be written as $\hat{H}^{(b)}$. The topologically nontrivial regime is shown in $\theta\in (-\pi/2,\pi/2)$. For the large size case, a pair of pure imaginary edge modes emerges under OBC regardless of the value of $\gamma$ \cite{Klett}. By applying transfer matrix method, we find in large $N$ limit, the nontrivial edge modes with energies $E_{L/R}^{(b)}=\pm i\gamma$ with the subscript '$L$' ('$R$') representing the left (right) mode emerge in the regime $\theta \in (-\pi/2,\pi/2)$ (see Appendix A). For $E_{L}^{(b)}=i\gamma$, the edge mode is localized at the left side with the wave function $\varphi_{j,B}(i\gamma)=0$ and $\varphi_{j,A}(i\gamma)=(-\xi)^{j-1}\varphi_{1,A}(i\gamma)$ for $j<N/2$ and the wave function of the right-side edge mode with the energy $E_{R}^{(b)}=-i\gamma$ is $\varphi_{j,A}(-i\gamma)=0$ and $\varphi_{j,B}(-i\gamma)=(-\xi)^{N-j}\varphi_{N,B}(-i\gamma)$ for $j>N/2$. Different from the breakup of the zero modes in a finite Hermitian lattice, one finds that the zero modes of the edge states can be recovered by non-Hermitian degeneracies through gain and loss modulations \cite{WangeSong}.

\begin{figure}[tbp]
	\begin{center}
		\includegraphics[width=.5 \textwidth] {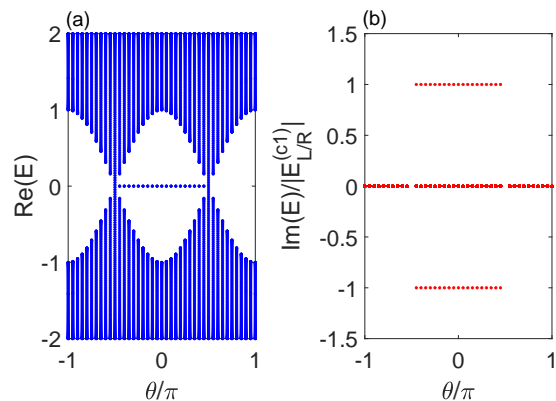}
	\end{center}
	\caption{(Color online) (a) Real part of the spectra of the SSH model with two additional conjugated imaginary potentials at both ends as the function of $\theta$. (b) Rescaling imaginary part of the spectra $\mathrm{Im}(E)/|E_{L/R}^{(c1)}|$ as the function of $\theta$. Here, $J=1$, $\Delta=0.5, \gamma=0.5$, and $N=100$ under OBC.}\label{Fig2}
\end{figure}

Next, we consider the $\mathcal{PT}$-symmetric SSH model with a pair of complex conjugate potentials described by Hamiltonian (\ref{eq1}) for case (c) as shown in Fig. \ref{Fig1}(c). When $n=1,\alpha^{\prime}=A$, i.e., $\hat{H}_1^{(c1)}=i\gamma\hat{c}_{1,A}^{\dagger}\hat{c}_{1,A}-i\gamma\hat{c}_{N,B}^{\dagger}\hat{c}_{N,B}$, it is the SSH model with a pair of the gain and loss acting at the two end sites which was numerically studied in \cite{BaogangZhu,Klett,Yuce}. In the topologically trivial regime, four imaginary energies' bound states emerge when $\gamma>\gamma_c$ with $\gamma_c$ being the critical point from $\mathcal{PT}$-symmetry unbroken to broken region, and the energies of the four bound states are calculated in Appendix A. Here, the bound states induced by the defects are located at the positions of the two conjugated imaginary defect potentials and have complex energies. The profiles of the bound states tend to zero when $|\tilde{j}-\tilde{j}_0| \to \infty$, i.e. $\lim_{|\tilde{j}-\tilde{j}_0|\to \infty}|\varphi_{\tilde{j}}|\to 0$ with $\tilde{j}_0$ being the position of the defect. The emergence of the bound states is caused by the imaginary defect potentials, which is similar to the Hermitian case with real defect potentials and the positions of the bound states depend on the defects. In the topologically nontrivial regime and large $N$ limit, there are two nontrivial edge modes with the imaginary energies (see Appendix A)
\begin{equation}\label{eq3}
E_{L/R}^{(c1)}=\pm\frac{(i\gamma+\frac{J_{+}^2}{i\gamma})+\sqrt{(i\gamma+\frac{J_{+}^2}{i\gamma})^2+4(J_{-}^2-J_{+}^2)}}{2},
\end{equation}
and $2N-2$ pure real bulk energies for arbitrary $\gamma\ne 0$ \cite{BaogangZhu} and no bound states exist. The phase diagram of case (c) with $n=1$, $\alpha^{\prime}=A$, and $\Delta=0.5$ is shown in Appendix C. Figure \ref{Fig2}(a) shows the real part of the spectrum of the SSH model with two additional conjugated imaginary potentials at both ends versus $\theta/\pi$ with $\Delta=0.5, \gamma=0.5$, and $N=100$. The topologically nontrivial regime characterized by exact-zero modes of the real part is influenced by the $\mathcal{PT}$-symmetric potentials under OBC. The nonzero imaginary energies of the edge modes scaled by $|E_{L/R}^{(c1)}|$ are exactly equal to $\pm 1$ [Fig. \ref{Fig2}(b)]. The nontrivial edge states in their real energy spectra are degenerate to an exact-zero state for the arbitrary potential amplitude $\gamma$ in the large $N$ limit.

\begin{figure}[tbp]
	\begin{center}
		\includegraphics[width=.5 \textwidth] {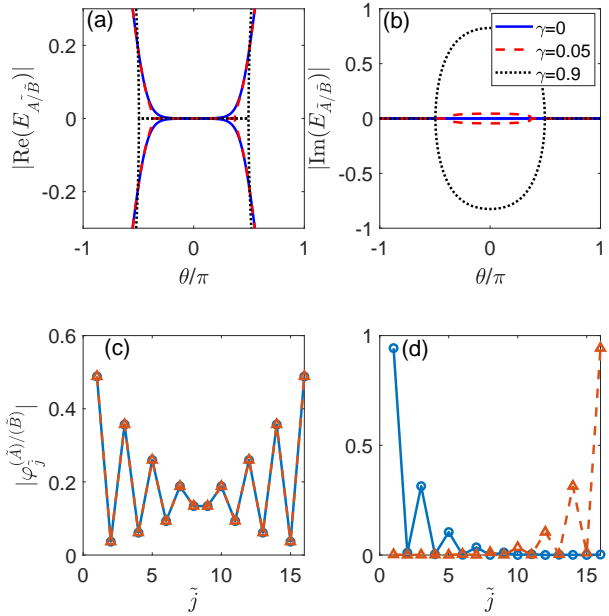}
	\end{center}
	\caption{(Color online) The (a) real and (b) imaginary  parts of two edge modes $E_{\tilde{A}}$ and $E_{\tilde{B}}$ of the SSH model with two additional conjugated imaginary potentials at both ends as a function of $\theta$ in non-Hermitian cases $\hat{H}^{(c1)}$ with $J=1$, $\Delta=0.5$, $N=8$, and different $\gamma$ in a finite system and profiles of the two edge modes $|\varphi_{\tilde{j}}^{(\tilde{A})}|$ and $|\varphi_{\tilde{j}}^{(\tilde{B})}|$ for (c) $\theta=0.4\pi$ and (d) $\theta=0$ with $J=1$, $\Delta=0.5$, $\gamma=0.05$, and $N=8$.}\label{Fig3}
\end{figure}

\begin{figure}[tbp]
	\begin{center}
		\includegraphics[width=.5 \textwidth] {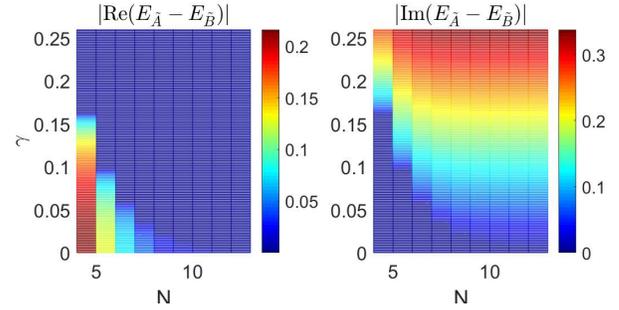}
	\end{center}
	\caption{(Color online) (Left) $|\mathrm{Re}(E_{\tilde{A}}-E_{\tilde{B}})|$ and (right) $|\mathrm{Im}(E_{\tilde{A}}-E_{\tilde{B}})|$ as functions of $N$ and $\gamma$ with $J=1$, $\Delta=0.5$, and $\theta=\pi/3$ for the model described by the Hamiltonian $\hat{H}^{(c1)}$.}\label{Fig4}
\end{figure}

We extend the above results and analyses to a finite system. Figures \ref{Fig3}(a) and \ref{Fig3}(b) show the real and imaginary parts of the mode spectra of two edge states with the energies $E_{\tilde{A}}$ and $E_{\tilde{B}}$ as the functions of $\theta$ in a small $N$ case ($N=8$) with $\Delta=0.5$ and different $\gamma$. The two edge modes in the Hermitian (blue curve) with purely real values are detuned from zero. Take $\theta=0$, $\gamma=0$, and $N=8$ as an example, $|E_{\tilde{A}}-E_{\tilde{B}}|\approx 4.0\times 10^{-4}$. As seen in Fig. \ref{Fig3}(a) for $\gamma=0.05$, two edge modes merge to exact-zero modes for the real part at $\theta\in(-0.371\pi,0.371\pi)$. With the increase of $\gamma$, we can see that the regime of the two edge modes in which their real parts tend to be zero enlarges and it finally reaches $\theta\in(-0.5\pi,0.5\pi)$ when $\gamma\approx 0.9$ [see Fig. \ref{Fig3}(a)]. Their imaginary parts split into two branches [see Fig. \ref{Fig3}(b)]. In Figs. \ref{Fig3}(c) and \ref{Fig3}(d), we show the profiles of the edge modes $|\varphi_{\tilde{j}}^{(\tilde{A})}|$ and $|\varphi_{\tilde{j}}^{(\tilde{B})}|$ with $\Delta=0.5$, $\gamma=0.05$, $N=8$, and different $\theta$. The real parts of the edge modes at $\theta=0.4\pi$ are nonzero values, whose profiles localize at both boundaries due to the boundary coupling. When $\theta=0$, the exact-zero modes present a single-boundary localized behavior due to the breaking of $\mathcal{PT}$ symmetry. To clearly demonstrate the recovery of zero modes, we plot diagrams of $|\mathrm{Re}(E_{\tilde{A}}-E_{\tilde{B}})|$ and $|\mathrm{Im}(E_{\tilde{A}}-E_{\tilde{B}})|$ as the functions of $N$ and $\gamma$ with $\Delta=0.5$ and $\theta=\pi/3$ shown in Fig. \ref{Fig4}. $|\mathrm{Re}(E_{\tilde{A}}-E_{\tilde{B}})|$ can tend to zero with increasing $N$ and $\gamma$ and the corresponding $|\mathrm{Im}(E_{\tilde{A}}-E_{\tilde{B}})|$ approaches to finite values. This indicates that one can recover the exact-zero modes by increase $N$ and $\gamma$ in such case. However, it is not clear whether or not the non-zero edge modes can always be recovered by the gain or loss strength in a finite system. To answer this question we consider the SSH model with $\mathcal{PT}$-symmetric potentials of other forms in the next section.

\section{Failure of recovery of zero modes}

\begin{figure}[tbp]
	\begin{center}
		\includegraphics[width=.5 \textwidth] {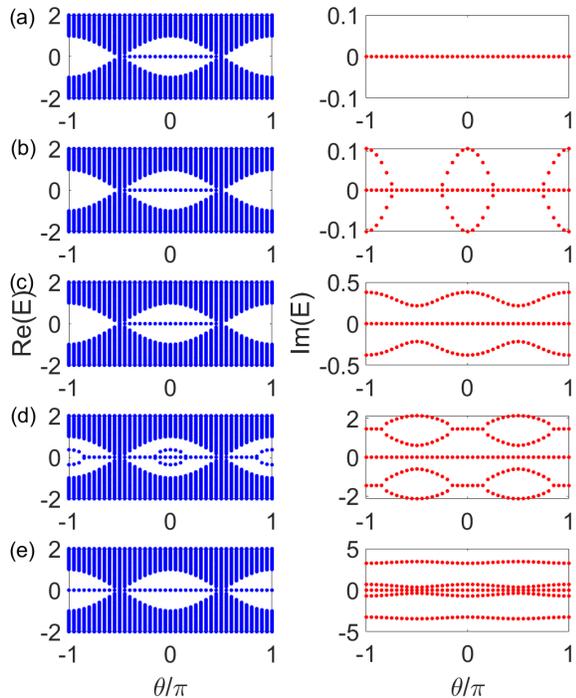}
	\end{center}
	\caption{(Color online) The real and imaginary parts of the spectra of the SSH model subjected to the non-Hermitian potentials described by $\hat{H}_1^{(c2)}=-i\gamma\hat{c}_{1,B}^{\dagger}\hat{c}_{1,B}+i\gamma\hat{c}_{N,A}^{\dagger}\hat{c}_{N,A}$ as a function of $\theta/\pi$ with $J=1$, $\Delta=0.5$, $N=50$, and different $\gamma$: (a) $\gamma=0.3$, (b) $\gamma=0.6$, (c) $\gamma=1$, (d) $\gamma=3$ and (e) $\gamma=4$ under OBC. The left column represents the real parts of the spectra and the right one shows the imaginary parts.}\label{Fig5}
\end{figure}

In this section, we consider case (c) for $n=1$ and $\alpha^{\prime}=B$ in large $N$ limit. Figure \ref{Fig5} shows the real and imaginary parts of the spectra of the SSH model subject to the potential described by $\hat{H}_1^{(c2)}=-i\gamma\hat{c}_{1,B}^{\dagger}\hat{c}_{1,B}+i\gamma\hat{c}_{N,A}^{\dagger}\hat{c}_{N,A}$ as a function of $\theta/\pi$ with $\Delta=0.5$, $N=50$, and different $\gamma$. In the large $N$ limit, the real parts of the spectra have a similar feature as that of standard SSH model when $\gamma<2.4$, i.e., there exist edge modes with $E=0$ in the topologically nontrivial regime. For the weak imaginary defect potentials $\gamma<0.475$ [Fig.\ref{Fig5} (a) for $\gamma=0.3$], the whole imaginary parts of the spectra tend to zero, even in the regime of $J_{-}<J_{+}$. When $\gamma\ge 0.475$, four complex energies of the bound states first emerge at $\theta=\pm \pi$. 
With the increase of $\gamma$, the regions of the $\mathcal{PT}$ unbroken phase shrink in both regimes which is shown in Fig. \ref{Fig5}(b) with $\gamma=0.6$. When $\gamma \ge 0.71$, the system totally immerses into the $\mathcal{PT}$-symmetry broken regime by rolling $\theta$ from $-\pi$ to $\pi$ as seen in Fig.\ref{Fig5}(c) with $\gamma=1$. An interesting observation is that when $\gamma\ge 2.4$, in addition to the zero edge modes in $\theta\in (-\pi/2,\pi/2)$, additional modes can be observed in the gap of the real parts. In Fig. \ref{Fig5}(d) with $\gamma=3$, the energies of the bound states become purely imaginary in $\theta\in (-0.84\pi,-0.16\pi)$ and $\theta\in (0.16\pi,0.84\pi)$. When $\gamma$ is larger than $3.08$, all bound states energies become purely imaginary for $\theta\in[-\pi,\pi]$ [see Fig. \ref{Fig5}(e) with $\gamma=4$]. All complex modes shown in Fig. \ref{Fig5} coincide with the analytic solutions in Appendix A. As a conclusion, the phase diagram of this case with $\Delta=0.5$ is shown in  Appendix C. According to our numerical and analytic results, in this large $N$ system, two degenerate edge modes always exist for arbitrary $\gamma$ in the topologically nontrivial regime and four bound states are shown in the $\mathcal{PT}$-symmetry broken regime with energies carrying nonzero imaginary part. The positions of the bound states depend on the gain and loss positions shown in Appendix B which have no relation to the topological property.

\begin{figure}[tbp]
	\begin{center}
		\includegraphics[width=.5 \textwidth] {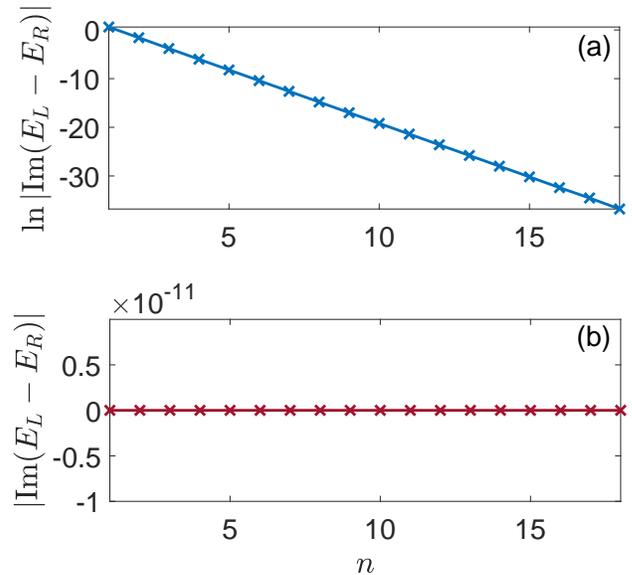}
	\end{center}
	\caption{(Color online)  $|\mathrm{Im}(E_{L}-E_{R})|$ as a function of $n$ with $J=1$, $N=36$, $\gamma=1$, $\Delta=0.5$, and $\theta=0$. Here, (a) $\alpha^{\prime}=A$ and (b) $\alpha^{\prime}=B$. }\label{Fig6}
\end{figure}

The nontrivial edge modes for the cases of arbitrary $n$ and $\alpha^{\prime}$ in the thermodynamic limit need more careful consideration. Figure \ref{Fig6} shows $|\mathrm{Im}(E_L-E_R)|$ as the function of $n$ for the SSH model subjected to one pair of $\mathcal{PT}$-symmetric defect potentials with different $\alpha^{\prime}$, $N=36$, $\gamma=1$, and $\Delta=0.5$ in the topologically nontrivial regime $\theta=0$. According to Fig. \ref{Fig6}(b) with $\alpha^{\prime}=B$ and $\theta=0$, the system possesses the $\mathcal{PT}$ symmetry for small gain or loss, in which case the energies of nontrivial edge modes are purely real. As seen in Appendix A for $\alpha^{\prime}=B$ cases, when $(-\xi)^{N-n}\to 0$ for $J_{-}<J_{+}$, the energies of topologically nontrivial edge modes are $E=0$ for arbitrary $\gamma$ and the wave functions of the edge modes on the two ends are $\psi_{L}(0)=[\varphi_{1,A}(0),0,-\xi \varphi_{1,A}(0),0,(-\xi)^{2}\varphi_{1,A}(0),0,\dots]^{T}$ and $\psi_{R}(0)=[\dots,0,(-\xi)^{2} \varphi_{N,B}(0),0,-\xi\varphi_{N,B}(0),0,\varphi_{N,B}(0)]^{T}$, respectively. However, for $\alpha^{\prime}=A$ as shown in Fig. \ref{Fig6}(a), the situation is changed. Two imaginary energies corresponding to edge modes emerge in the small $n$ case, and with the increase of $n$, $|\mathrm{Im}(E_L-E_R)|\propto e^{-n/0.455}$, sharing an exponential decay for $\theta=0$ and $\Delta=0.5$. The authors of \cite{JinLiang} gave the critical point $\gamma_c$ from $\mathcal{PT}$-symmetry unbroken to broken region in the thermodynamic limit. For $\alpha^{\prime}=A$, the minimum $\gamma_c$ in region $\theta\in [-\pi/2,\pi/2]$ tends to zero in small $n$, and grows with the increase of $n$, while for $\alpha^{\prime}=B$, the minimum $\gamma_c$ is finite which corresponds to our results and the phase diagrams of case (c) with small $n$ shown in Appendix C. It concludes that the energies of the two nontrivial edge modes for $\alpha^{\prime}=B$ case always amount to exact-zero for an arbitrary value of $\gamma$ in the large $N$ limit and for $\alpha^{\prime}=A$, our numerical calculation elucidates that in the small $n$ case the balanced gain and loss can strongly affect the energies of the edge modes. 

\begin{figure}[tbp]
	\begin{center}
		\includegraphics[width=.5 \textwidth] {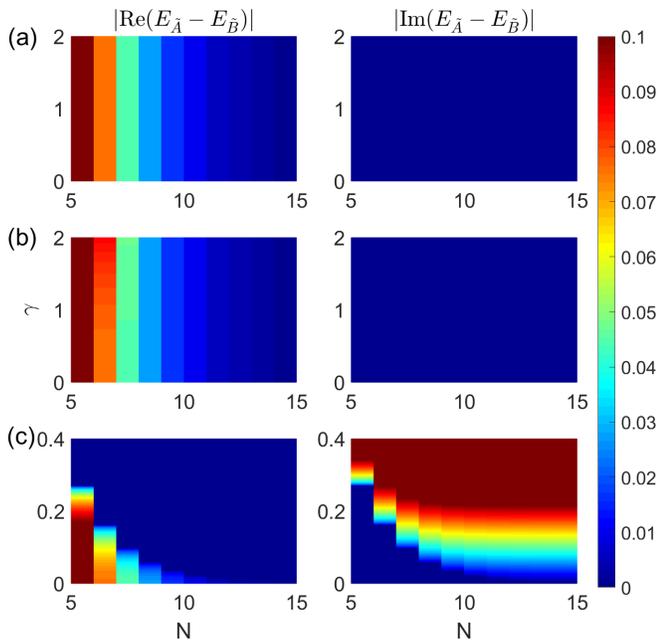}
	\end{center}
	\caption{(Color online)  (Left column) $|\mathrm{Re}(E_{\tilde{A}}-E_{\tilde{B}})|$ and (right column) $|\mathrm{Im}(E_{\tilde{A}}-E_{\tilde{B}})|$ as functions of $N$ and $\gamma$ with $J=1$, $\Delta=0.5$ and $\theta=\pi/3$ for $\hat{H}^{(c)}$ with (a) $n=1$, $\alpha^{\prime}=B$, (b) $n=2$, $\alpha^{\prime}=B$, and (c) $n=2$, $\alpha^{\prime}=A$, respectively.}\label{Fig7}
\end{figure}

We calculate $|\mathrm{Re}(E_{\tilde{A}}-E_{\tilde{B}})|$ and $|\mathrm{Im}(E_{\tilde{A}}-E_{\tilde{B}})|$ as the functions of system size $N$ and the gain or loss strength $\gamma$ to reveal the recovery of zero modes in Fig. \ref{Fig7} with $\Delta=0.5$, $\theta=\pi/3$ for different $n$ and $\alpha^{\prime}$. For $n=1$ and $\alpha^{\prime}=B$ shown in Fig. \ref{Fig7}(a), the values of $|\mathrm{Re}(E_{\tilde{A}}-E_{\tilde{B}})|$ are finite for arbitrary $\gamma$ in the small $N$ limit and the imaginary part of $E_{\tilde{A}}-E_{\tilde{B}}$ always approaches zero. One can recover the edge modes only by increasing the system size $N$ in such a case. We find that Fig. \ref{Fig7}(b) with $n=2$ and $\alpha^{\prime}=B$ exhibits similar behaviors as shown in Fig. \ref{Fig7}(a). The recovery of zero modes by adding the $\mathcal{PT}$-symmetric potential at $\alpha^{\prime}=B$ is unsuccessful in the small $N$ limit. However, when we locate two additional conjugated imaginary on-site potentials at $\alpha^{\prime}=A$, the situation returns to the case (b). As shown in Fig. \ref{Fig7}(c) with $n=2$ and $\alpha^{\prime}=A$, $|\mathrm{Re}(E_{\tilde{A}}-E_{\tilde{B}})|$ can recover to zero and $|\mathrm{Im}(E_{\tilde{A}}-E_{\tilde{B}})|$ breaks to finite values by increasing either $N$ or $\gamma$. This implies that the recovery of exact-zero modes by increasing $\gamma$ depends on the type of $\mathcal{PT}$-symmetric potential. If the zero modes are free from the influence of the additional $\mathcal{PT}$-symmetric potentials in large $N$ limit, the recovery shall be failed by only increasing $\gamma$.

\section{Edge modes in a spin-orbit-coupled SSH chain with $\mathcal{PT}$ symmetry}

\begin{figure}[tbp]
	\begin{center}
		\includegraphics[width=.5 \textwidth] {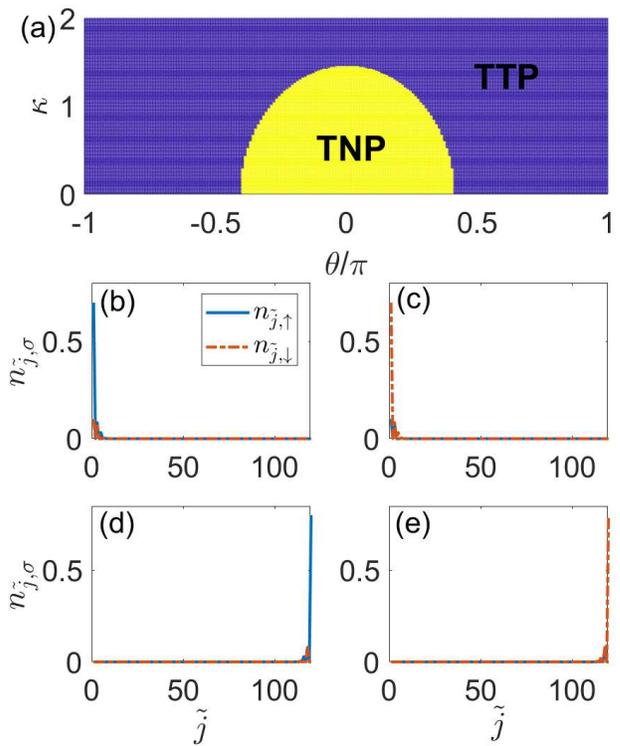}
	\end{center}
	\caption{(Color online) (a) Topological phase diagram of Hamiltonian (\ref{eq4}) as the function of $\kappa$ and $\theta$ with $J=1$, $\Delta=0.5$. (b) - (e) Density distributions $n_{\tilde{j},\sigma}$ of the four edge modes with $N=60$, $J=1$, $\Delta=0.5$, $\theta=0$, and $\kappa=0.5$ under OBC.}\label{Fig8}
\end{figure}

To further clarify our conclusions, we consider a toy model which is a spin-orbit coupled SSH chain with Hamiltonian
\begin{equation}\label{eq4}
\hat{H}_{\mathrm{SOC}}=\sum_j (\hat{\psi}^{\dagger}_{j,A}\mathcal{R}^{(1)}\hat{\psi}_{j,B}+\hat{\psi}^{\dagger}_{j,B}\mathcal{R}^{(2)}\hat{\psi}_{j+1,A} +H.c.),
\end{equation}
where $\hat{\psi}_{j,\alpha}^{\dagger}=(\hat{c}_{j,\alpha,\uparrow}^{\dagger},\hat{c}_{j,\alpha,\downarrow}^{\dagger})$ and $\hat{c}_{j,\alpha,\sigma}^{\dagger}$ creates a fermion with spin $\sigma=\uparrow,\downarrow$ in the $j$-th unit cell. The hopping matrices have the form $\mathcal{R}^{(1)}=\pm i( J_{-}\sigma_z+\kappa \sigma_y)$ and $\mathcal{R}^{(2)}=J_{+}I\pm i \kappa \sigma_y$, along the $\pm \hat{x}$ direction, respectively where $\kappa$ controls the spin-orbit coupling strength, $\sigma_{y(z)}$ is the Pauli matrix of the $y(z)$ component and $I$ is the $2\times 2$ identity matrix. The diagonal terms of $\mathcal{R}$ describe spin-conserving hopping and the off-diagonal terms describe spin-flipping hopping between the nearest neighbors. To simplify the discussion, we set $J=1$ and $\Delta=0.5$. Figure \ref{Fig8}(a) shows the topological phase diagram of the Hamiltonian (\ref{eq4}) as the function of $\theta$ and $\kappa$. Phase "TNP" denotes the topologically nontrivial phase with the winding number $W=2$ where four zero edge modes can be detected, while phase "TTP" is the topologically trivial regime with $W=0$. We can find that with the increase of $\kappa$, the topologically nontrivial regime shrinks. Figures \ref{Fig8}(b) to \ref{Fig8}(e) show the density distributions $n_{\tilde{j},\sigma}$ of the four edge modes with $N=60$, $J=1$, $\Delta=0.5$, $\theta=0$, and small $\kappa=0.5$ under OBC. We mark the edge modes localized at the left boundary as $A_1$ and $A_2$ modes in Figs. \ref{Fig8}(b) and \ref{Fig8}(c), respectively, and the $B_1$ and $B_2$ modes are localized at the right boundary, which is shown, respectively, in Figs. \ref{Fig8}(d) and \ref{Fig8}(e). For the $A_1$ ($B_1$) mode, the profile of spin-$\uparrow$($\downarrow$) at $\tilde{j}=1$ is dominant. And for the $A_2$ ($B_2$) mode, $|\varphi_{2N,\uparrow}^{A_{2}}|\gg |\varphi_{2N,\downarrow}^{A_{2}}|$ ($|\varphi_{2N,\uparrow}^{B_{2}}|\ll|\varphi_{2N,\downarrow}^{B_{2}}|$) in the large $N$ limit. However, when the system size becomes finite, the four edge modes deviate from exact-zero energy with two-fold degeneracy and the two-fold degenerate edge modes are coupled together.

To study the recovery of the edge modes in the small $N$ limit, additional $\mathcal{PT}$-symmetric potentials are considered, and the total Hamiltonian can be described as
$\hat{H}_{\mathrm{SOC}}^{(\beta)}=\hat{H}_{\mathrm{SOC}}+\hat{H}_{2}^{(\beta)}$, where $\hat{H}_{2}^{(\beta)}$ is the $\mathcal{PT}$-symmetric non-Hermitian term. In this section, three cases are considered as follows: (a) $\hat{H}_{2}^{(a)}=i\gamma(\hat{c}_{1,A,\uparrow}^{\dagger}\hat{c}_{1,A,\uparrow}-\hat{c}_{N,B,\downarrow}^{\dagger}\hat{c}_{N,B,\downarrow})$, (b) $\hat{H}_{2}^{(b)}=i\gamma(\hat{c}_{1,A,\downarrow}^{\dagger}\hat{c}_{1,A,\downarrow}-\hat{c}_{N,B,\uparrow}^{\dagger}\hat{c}_{N,B,\uparrow})$, and (c) $\hat{H}_{2}^{(c)}=i\gamma\sum_{\sigma}(-1)^{\tilde{\sigma}}(\hat{c}_{1,A,\sigma}^{\dagger}\hat{c}_{1,A,\sigma}-\hat{c}_{N,B,\sigma}^{\dagger}\hat{c}_{N,B,\sigma})$, where $\tilde{\sigma}=1$($2$) for $\sigma=\uparrow$($\downarrow$). The additional $\mathcal{PT}$-symmetric term is not expected to alter the topological phase of such model. Taking $J=1$, $\Delta=0.5$, $\theta=0$, small $\kappa=0.5$, and $\gamma=1$ as an example, we numerically calculate the energies of the four edge modes of the three cases. For case (a) proposed in this section in the large $N$ limit, only modes $A_1$ and $B_2$ are influenced by $\hat{H}_{2}^{(a)}$ and the energies of modes $A_2$ and $B_1$ are exact-zero. When we turn to a finite system size, taking $N=5$ as an example, $|\mathrm{Re}(E_{A_1}-E_{B_2})|$ can be recovered to zero and $|\mathrm{Re}(E_{A_2}-E_{B_1})|=0.0256$ for $\gamma=1$. When $\hat{H}_{2}^{(b)}$ term is considered, the energies of modes $A_2$ and $B_1$ become imaginary and modes $A_1$ and $B_2$ keep zero in the large $N$ case. Correspondingly, for a small $N$ by increasing $\gamma$, modes $A_2$ and $B_1$ can be recovered and the recovery of modes $A_1$ and $B_2$ breaks down. For the infinite system with the Hamiltonian $\hat{H}_{\mathrm{SOC}}^{(c)}$, we find that all the four edge modes break the $\mathcal{PT}$ symmetry and become the imaginary energy states. In the small size case, we can recover all the edge modes by increasing the gain or loss strength $\gamma$ of $\hat{H}_{\mathrm{SOC}}^{(c)}$. Our results indicate that if balanced gain and loss are localized on the sites with nonzero distributions for the edge modes, the edge states will break the $\mathcal{PT}$ symmetry which can induce the recovery of the real parts of the edge modes to exact-zero value in small system size.

\section{Conclusion}

In conclusion, we studied the fate of zero modes of $\mathcal{PT}$-symmetric SSH models subject to the non-Hermitian on-site potentials of different forms in large and small $N$ limits, respectively. The topologically nontrivial zero modes will deviate from exact-zero energy in a finite system and some types of $\mathcal{PT}$-symmetric potentials may recover the near-zero modes to exact-zero ones, which is closely related to the results in large $N$ limits. For a standard SSH model, the probabilities of edge modes are staggered decreasing from the boundaries, and the distributions at sublattice $B$ ($A$) for the left (right) edge mode are zero. The influence of the gain and loss on the edge states is remarkable for small $n$ and $\alpha^{\prime}=A$. By adding $\mathcal{PT}$-symmetric potentials on the sites with the nonvanishing distribution probabilities, it will lead to $\mathcal{PT}$ symmetry break down \cite{JinLiang,Bendix}. When the edge modes break the $\mathcal{PT}$ symmetry, the energies of edge states become a conjugate imaginary pair in both large and small system size limits. However, when the $\mathcal{PT}$-symmetric potentials are localized on the sites with vanishing distribution probabilities of the edge modes, the recovery of the zero modes break down by modulating $\gamma$ in a finite system since the edge states are hardly affected by the $\mathcal{PT}$-symmetric potentials. Our results can be easily simulated in a silicon waveguide platform with controlled gain or loss.

\begin{acknowledgements}
Z. Xu is supported by the NSF of China under Grant No. 11604188 and STIP of Higher Education Institutions in Shanxi under Grant No. 2019L0097. S. Chen was supported by the NSFC (Grant No. 11974413) and the NKRDP of China (Grants No. 2016YFA0300600 and No. 2016YFA0302104).  L.B. Fu is supported by the National Natural Science Foundation of China (Grants No. 11725417, No. 11575027, and No. U1930403), and Science Challenge Project (Grant No. TZ2018005). Y. Zhang is supported by NSF of China under Grant No. 11674201. This work is also supported by NSF for Shanxi Province Grant No.1331KSC.
	
\end{acknowledgements}

\section*{Appendix A: Solutions of the Edge modes and bound states in infinite systems}

In this Appendix, we solve the spectrum for the cases $\hat{H}^{(b,c)}$ shown in Figs. \ref{Fig1}(b) and \ref{Fig1}(c) in the main text. For the non-Hermitian part of the Hamiltonian described as $\hat{H}_1^{(b)}=\sum_{j}(i\gamma\hat{c}^{\dagger}_{j,A}\hat{c}_{j,A}-i\gamma\hat{c}^{\dagger}_{j,B}\hat{c}_{j,B})$ which can keep the total Hamiltonian behaving as a $\mathcal{PT}$ symmetry under both OBCs and PBCs, we can employ the transfer matrix method to gain insight into the properties of the edge modes of the system under OBCs. For a single-particle state $\psi(E)=\sum_{j,\alpha}\varphi_{j,\alpha}(E)\hat{c}_{j,\alpha}^{\dagger}|0\rangle$ with an elementary excitation energy $E$, we can solve the Schr\"{o}dinger equation and obtain the difference equations as follows:
\begin{align}\label{Seq9}
(E-i\gamma)\varphi_{1,A} &= J_{-} \varphi_{1,B}, \notag\\
(E-i\gamma)\varphi_{j,A} &= J_{-} \varphi_{j,B}+J_{+}\varphi_{j-1,B} \quad j\in(1,N], \notag \\
(E+i\gamma)\varphi_{j,B} &= J_{-} \varphi_{j,A}+J_{+}\varphi_{j+1,A} \quad j\in[1,N), \notag \\
(E+i\gamma)\varphi_{N,B} &= J_{-} \varphi_{N,A}. \notag\\
\end{align}
In the large $N$ limit, the energies of the edge modes are $E_{L/R}^{(b)}=\pm i\gamma$ in the regime $J_{-}<J_{+}$ where $L$ ($R$) represents the left (right) mode.
In the topologically nontrivial regime, for $E_{L}^{(b)}=i\gamma$, the edge mode localized at the left side with $\varphi_{j,B}=0$ and $\varphi_{j,A}=(-J_{-}/J_{+})^{j-1}\varphi_{1,A}$. For the right-side edge mode, $E_{R}^{(b)}=-i\gamma$ and the wave function is $\varphi_{j,A}=0$ and $\varphi_{j,B}=(-J_{-}/J_{+})^{N-j}\varphi_{N,B}$.

The SSH model with the $\mathcal{PT}$-symmetric potentials $\hat{H}_1^{(c)}$ is shown in Fig.\ref{Fig1}(c) in the main text. When $n=1, \alpha^{\prime}=A$, it shows a pair of imaginary potential at two ends,  $\hat{H}_1^{(c1)}=i\gamma\hat{c}^{\dagger}_{1,A}\hat{c}_{1,A}-i\gamma\hat{c}^{\dagger}_{N,B}\hat{c}_{N,B}$.  We construct the transfer matrix representation of the recursion relation as follows:
\begin{align}\label{Seq1}
(E-i\gamma)\varphi_{1,A} &= J_{-} \varphi_{1,B}, \notag\\
E\varphi_{j,A} &= J_{-} \varphi_{j,B}+J_{+}\varphi_{j-1,B} \quad j\in(1,N], \notag \\
E\varphi_{j,B} &= J_{-} \varphi_{j,A}+J_{+}\varphi_{j+1,A} \quad j\in[1,N), \notag \\
(E+i\gamma)\varphi_{N,B} &= J_{-} \varphi_{N,A}. \notag\\
\end{align}
$E=0$ no longer represents the edge modes in Eq. (\ref{Seq1}). We eliminate $A$ and $B$ sublattices in Eq. (\ref{Seq1}), respectively, and obtain
\begin{align}\label{Seq2}
&(E^2-i\gamma E-J_{-}^2)\varphi_{1,A} = J_{-}J_{+}\varphi_{2,A}, \notag\\
&(E^2-J_{-}^2-J_{+}^2)\varphi_{j,A} = J_{-}J_{+} (\varphi_{j-1,A}+\varphi_{j+1,A}) \quad j\in(1,N), \notag \\
&(E^2-\frac{E}{E+i\gamma}J_{-}^2-J_{+}^2)\varphi_{N,A}=J_{-}J_{+}\varphi_{N-1,A} \notag \\
&(E^2-\frac{E}{E-i\gamma}J_{-}^2-J_{+}^2)\varphi_{1,B}=J_{-}J_{+}\varphi_{2,B} \notag \\
&(E^2-J_{-}^2-J_{+}^2)\varphi_{j,B} = J_{-}J_{+} (\varphi_{j-1,B}+\varphi_{j+1,B}) \quad j\in(1,N), \notag \\
&(E^2+i\gamma E-J_{-}^2)\varphi_{N,B} = J_{-}J_{+} \varphi_{N-1,B}. \notag\\
\end{align}
Equation (\ref{Seq2}) can be written into matrix form as follows:
\begin{align}\label{Seq3}
\begin{pmatrix}
\varphi_{2,A} \\
\varphi_{1,A} \\
\end{pmatrix} &=T_1\begin{pmatrix}
\varphi_{1,A} \\
\varphi_{0,A} \\
\end{pmatrix}, \notag \\
\begin{pmatrix}
\varphi_{j+1,A} \\
\varphi_{j,A} \\
\end{pmatrix} &=T\begin{pmatrix}
\varphi_{j,A} \\
\varphi_{j-1,A} \\
\end{pmatrix} \quad j\in(1,N), \notag \\
\begin{pmatrix}
\varphi_{N-1,A} \\
\varphi_{N,A} \\
\end{pmatrix} &=T_2\begin{pmatrix}
\varphi_{N,A} \\
\varphi_{N+1,A} \\
\end{pmatrix}, \notag \\
\begin{pmatrix}
\varphi_{2,B} \\
\varphi_{1,B} \\
\end{pmatrix} &=T_3\begin{pmatrix}
\varphi_{1,B} \\
\varphi_{0,B} \\
\end{pmatrix},\notag \\
\begin{pmatrix}
\varphi_{j-1,B} \\
\varphi_{j,B} \\
\end{pmatrix} &=T\begin{pmatrix}
\varphi_{j,B} \\
\varphi_{j+1,B} \\
\end{pmatrix} \quad j\in(1,N), \notag \\
\begin{pmatrix}
\varphi_{N-1,B} \\
\varphi_{N,B} \\
\end{pmatrix} &=T_4\begin{pmatrix}
\varphi_{N,B} \\
\varphi_{N+1,B} \\
\end{pmatrix}, \notag \\
\end{align}
where
\begin{align}\label{Seq4}
&T=\begin{pmatrix} \mu & -1 \\1 & 0 \end{pmatrix},\quad T_1=\begin{pmatrix} \nu_1 & -1 \\1 & 0 \end{pmatrix}, \quad T_2=\begin{pmatrix} \xi_2 & -1 \\1 & 0 \end{pmatrix}, \notag \\
&T_3=\begin{pmatrix} \xi_1 & -1 \\1 & 0 \end{pmatrix},\quad T_4=\begin{pmatrix} \nu_2 & -1 \\1 & 0 \end{pmatrix}, \notag \\
\end{align}
with $\mu=(E^2-J_{-}^2-J_{+}^2)/J_{-}J_{+}$, $\nu_1=(E^2-i\gamma E-J_{-}^2)/J_{-}J_{+}$ , $\nu_2=(E^2+i\gamma E-J_{-}^2)/J_{-}J_{+}$, $\xi_1=(E^2-\frac{E}{E-i\gamma}J_{-}^2-J_{+}^2)/J_{-}J_{+}$, and $\xi_2=(E^2-\frac{E}{E+i\gamma}J_{-}^2-J_{+}^2)/J_{-}J_{+}$. We can diagonalize the transfer matrix $D=U^{-1}TU$ with
\begin{align}\label{Seq5}
D=\begin{pmatrix} \lambda_{-} & 0 \\ 0 & \lambda_{+}  \end{pmatrix},\quad U=\begin{pmatrix} \lambda_{-} & \lambda_{+} \\1 & 1 \end{pmatrix}, \notag \\ U^{-1}=\frac{1}{\sqrt{\mu^2-4}}\begin{pmatrix} -1 & \lambda_{+} \\1 & -\lambda_{-} \end{pmatrix}, \notag \\
\end{align}
and $\lambda_{\pm}=(\mu\pm\sqrt{\mu^2-4})/2$. First, we consider the left edge mode and the amplitudes of the wave function near the right side tends to zero in the large $N$ limit. Under OBCs with $\varphi_{0,A}=\varphi_{0,B}=0$, we can rewrite Eq. (\ref{Seq3}) as
\begin{align}\label{Seq6}
\begin{pmatrix} \varphi_{j+1,A} \\ \varphi_{j,A} \end{pmatrix} &= \frac{1}{\sqrt{\mu^2-4}} U \begin{pmatrix} (\lambda_{+}-\nu_1)\lambda^{j-1}_{-} \\  (\nu_1-\lambda_{-})\lambda^{j-1}_{+} \end{pmatrix}\varphi_{1,A}, \notag \\
\begin{pmatrix} \varphi_{j+1,B} \\ \varphi_{j,B} \end{pmatrix} &= \frac{1}{\sqrt{\mu^2-4}} U \begin{pmatrix} (\lambda_{+}-\xi_1)\lambda^{j-1}_{-} \\  (\xi_1-\lambda_{-})\lambda^{j-1}_{+} \end{pmatrix}\frac{E-i\gamma}{J_{-}}\varphi_{1,A}. \notag \\
\end{align}
The necessary conditions for the existence of the edge modes in the topologically nontrivial regime are $|\lambda_{-}|>1$, $\lambda_{+}=\nu_{1}$ and $\lambda_{+}=\xi_{1}$ for the localized left edge. The energy of the left edge mode is
\begin{equation}\label{Seq7}
E_{L}^{(c1)}=\frac{(i\gamma+\frac{J_{+}^2}{i\gamma})+\sqrt{(i\gamma+\frac{J_{+}^2}{i\gamma})^2+4(J_{-}^2-J_{+}^2)}}{2}.
\end{equation}
We can also obtain the energy of the right edge mode
\begin{equation}\label{Seq8}
E_{R}^{(c1)}=-\frac{(i\gamma+\frac{J_{+}^2}{i\gamma})+\sqrt{(i\gamma+\frac{J_{+}^2}{i\gamma})^2+4(J_{-}^2-J_{+}^2)}}{2}.
\end{equation}
However, in the topologically trivial regime, when $\gamma>\gamma_c$ with $\gamma_c$ being the transition point from the $\mathcal{PT}$-symmetric region to the spontaneously broken $\mathcal{PT}$-symmetry region, there exists four bound states with energies
\begin{equation}\label{Seq17}
E_{BS}^{(c1)}=\pm\frac{(i\gamma+\frac{J_{+}^2}{i\gamma})\pm\sqrt{(i\gamma+\frac{J_{+}^2}{i\gamma})^2+4(J_{-}^2-J_{+}^2)}}{2}.
\end{equation}

For $\alpha^{\prime}=B$, the transfer matrix method is applied with an elementary excitation energy $E$ and the difference equations as follows:
\begin{align}\label{Seq10}
&E\varphi_{1,A} = J_{-} \varphi_{1,B}, \notag\\
&E\varphi_{j,A} = J_{-} \varphi_{j,B}+J_{+}\varphi_{j-1,B} \quad j\in(1,N] ,j\ne n, \notag \\
&E\varphi_{j,B} = J_{-} \varphi_{j,A}+J_{+}\varphi_{j+1,A} \quad j\in[1,N) ,j\ne N+1-n, \notag \\
&(E+i\gamma)\varphi_{n,B} = J_{-} \varphi_{n,A}+J_{+}\varphi_{n+1,A}, \notag \\
&(E-i\gamma)\varphi_{N+1-n,A} = J_{-} \varphi_{N+1-n,B}+J_{+}\varphi_{N-n,B}, \notag \\
&E\varphi_{N,B} = J_{-} \varphi_{N,A}. \notag\\
\end{align}
We consider $N$ is large enough to make $(-\xi)^{N-n}\varphi_{1,A}\to 0$ and $(-\xi)^{N-n}\varphi_{N,B}\to 0$ in the topologically nontrivial regime  with $\xi=J_{-}/J_{+}$, so the energies of the edge modes $E=0$. The wave function of the left edge mode is $\psi_{L}(E=0)=[\varphi_{1,A},0,-\xi \varphi_{1,A},0,(-\xi)^{2}\varphi_{1,A},0,\dots]^{T}$, and the wave function of the right edge mode can be written as $\psi_{R}(E=0)=[\dots,0,(-\xi)^{2} \varphi_{N,B},0,-\xi\varphi_{N,B},0,\varphi_{N,B}]^{T}$. It indicates that for $\alpha^{\prime}=B$, the energies of the nontrivial edge modes approach to zero in the large $N$ limit. We take $n=1$, $\alpha^{\prime}=B$ as a concrete example. In the large $N$ case, $(-\xi)^{N-1}\varphi_{1,A}\to 0$ and $(-\xi)^{N-1}\varphi_{L,B}\to 0$ for $J_{-}<J_{+}$, thus the zero edge modes can be detected in the system. With the increase of $\gamma$, additional complex modes emerge corresponding to the bound states for $E\ne 0$. The equations for $A$ sublattice become
\begin{align}\label{Seq11}
\nu_1\varphi_{1,A} &= \varphi_{2,A}, \notag \\
\xi_3\varphi_{2,A}&=\varphi_{3,A}+\frac{E}{E+i\gamma}\varphi_{1,A}, \notag \\
\mu\varphi_{j,A} &=  \varphi_{j-1,A}+\varphi_{j+1,A} \quad j\in (2,N), \notag \\
\chi_2 \varphi_{N,A} &=  \varphi_{N-1,A}, \notag \\
\end{align}
and for $B$ sublattice are
\begin{align}\label{Seq12}
\chi_1\varphi_{1,B} &= \varphi_{2,B}, \notag \\
\mu\varphi_{j,B} &=  \varphi_{j-1,B}+\varphi_{j+1,B} \quad j\in (2,N), \notag \\
\xi_4\varphi_{N-1,B}&=\varphi_{N-2,A}+\frac{E}{E-i\gamma}\varphi_{L,B}, \notag \\
\nu_2 \varphi_{N,B} &=  \varphi_{N-1,B}, \notag \\
\end{align}
where $\xi_3=(E^2-J_{-}^2-\frac{E}{E+i\gamma}J_{+}^2)/J_{-}J_{+}$, $\xi_4=(E^2-J_{-}^2-\frac{E}{E-i\gamma}J_{+}^2)/J_{-}J_{+}$, $\chi_1=(E^2+i\gamma E-J_{-}^2-J_{+}^2)/J_{-}J_{+}$ and $\chi_2=(E^2-i\gamma E-J_{-}^2-J_{+}^2)/J_{-}J_{+}$.
We can follow the same procedure as the case shown in Fig. \ref{Fig1}(b). For the bound states near the left side, the necessary condition is
\begin{equation}\label{Seq13}
|\lambda_{-}|>1,\quad \lambda_{+}=\chi_1 ,\quad \lambda_{+}\nu_1=\xi_3\nu_1-\frac{E}{E+i\gamma}.
\end{equation}
Hence, the energies of the left bound states can be obtained by
\begin{equation}\label{Seq14}
E^3+i\gamma E^2-(J_{-}^2+J_{+}^2)E+\frac{J_{-}^2J_{+}^2}{i\gamma}=0
\end{equation}
and the roots should satisfy the necessary condition Eq. (\ref{Seq13}). For the bound states near the right side, the necessary condition is
\begin{equation}\label{Seq15}
|\lambda_{-}|>1,\quad \lambda_{+}=\chi_2 ,\quad \lambda_{+}\nu_2=\xi_4\nu_2-\frac{E}{E-i\gamma}.
\end{equation}
The energies of the right-side bound states satisfy
\begin{equation}\label{Seq16}
E^3-i\gamma E^2-(J_{-}^2+J_{+}^2)E-\frac{J_{-}^2J_{+}^2}{i\gamma}=0
\end{equation}
and also the necessary condition Eq. (\ref{Seq15}).

\section*{Appendix B: Bound states for case (c)}

\begin{figure}[tbp]
	\begin{center}
		\includegraphics[width=.5 \textwidth] {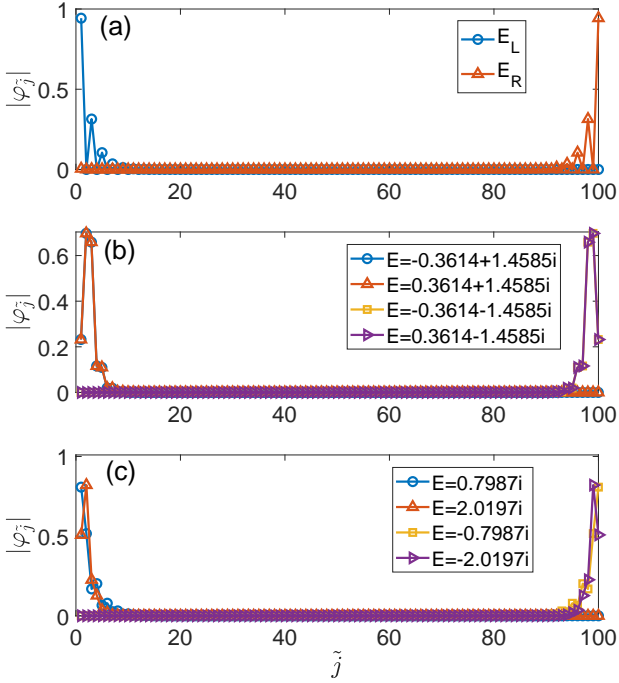}
	\end{center}
	\caption{(Color online) Profiles of (a) the zero modes and (b, c) the bound states with $J=1$, $\gamma=3$, $N=50$, and different $\theta$ under OBC. (a, b) $\theta=0$, (c) $\theta=2\pi/3$. Here, $n=1$ and $\alpha^{\prime}=B$.}\label{sFig1}
\end{figure}

\begin{figure}[tbp]
	\begin{center}
		\includegraphics[width=.5 \textwidth] {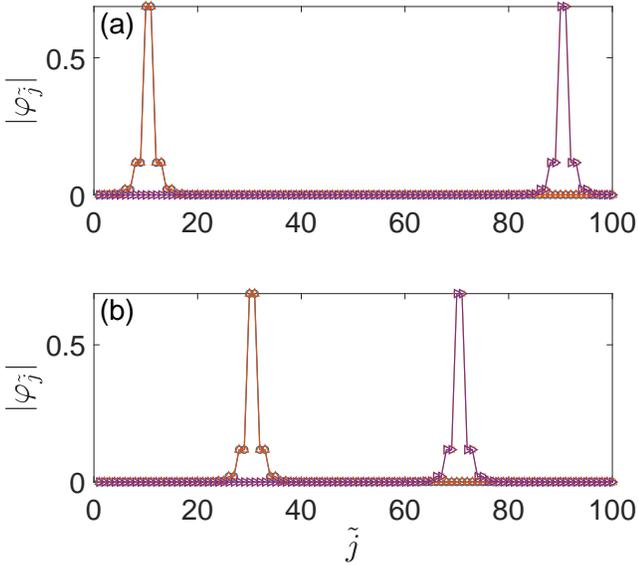}
	\end{center}
	\caption{(Color online) Profiles of the bound states with $J=1$, $\theta=0$, $\gamma=3$, $N=50$ under OBC. (a) $n=6$, $\alpha^{\prime}=A$ and (b) $n=15$, $\alpha^{\prime}=B$.}\label{sFig2}
\end{figure}

In this section, we consider the profiles of the bound states for $\hat{H}_1^{(c)}$. Firstly, we consider $n=1$, $\alpha^{\prime}=B$ case. As seen in the main text, when $\gamma\ge 0.475$, the bound states with complex energies emerge. Take $\gamma=3$ as an example. Fig.\ref{sFig1} shows the profiles of the zero modes and the bound states with $\gamma=3$, $N=50$ and different $\theta$ under OBC. For $\theta=0$, two degenerate zero modes are localized at the left and right boundary of the lattice shown in Fig.\ref{sFig1}(a). Four additional bound states are shown in Fig.\ref{sFig1}(b) with the energies being complex. In topological trivial regime shown in Fig.\ref{sFig1}(c) for $\theta=2\pi/3$, the number of the bound states with imaginary energies remains four but the topologically nontrivial zero modes disappear in this regime. Fig. \ref{sFig2} shows the profiles of the bound states with $\theta=0$, $\gamma=3$, $N=50$ and different $n$ and $\alpha^{\prime}$ under OBC. There are four bound states  in both Fig. \ref{sFig2}(a) and (b). The positions of the peaks of the bound states are determined totally by the positions of the gain and loss which is quite different from case of edge modes.

\section*{Appendix C: Phase diagrams of case (c) for small $n$}

\begin{figure}[tbp]
	\begin{center}
		\includegraphics[width=.5 \textwidth] {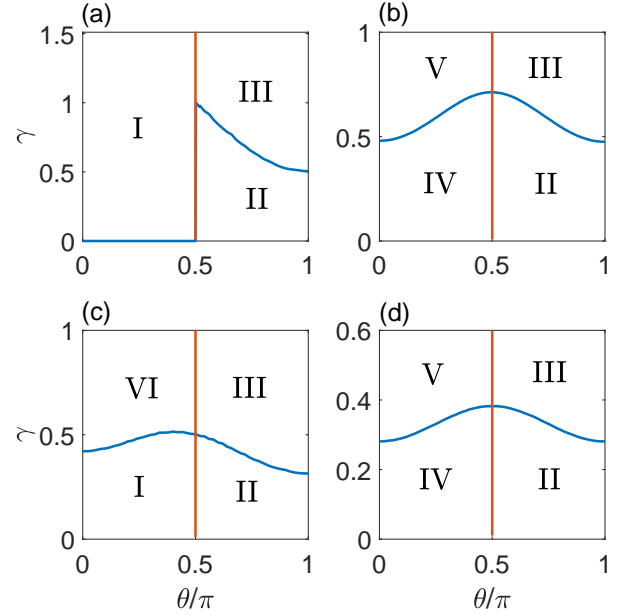}
	\end{center}
	\caption{(Color online) Phase diagrams as the functions of $\gamma$ and $\theta$ with $J=1$, $\Delta=0.5$, and different $n$ and $\alpha^{\prime}$. (a) $n=1$, $\alpha^{\prime}=A$, (b) $n=1$, $\alpha^{\prime}=B$, (c) $n=2$, $\alpha^{\prime}=A$, and (d) $n=2$, $\alpha^{\prime}=B$.}\label{sFig3}
\end{figure}

We further present here the $\gamma$-$\theta$ phase diagram of case (c) for different $\alpha^{\prime}$ in the small $n$ limit and the results are symmetric with respect to $\theta=0$. As shown in Fig. \ref{sFig3} with $J=1$ and $\Delta=0.5$, six phases emerge. For phase $\mathrm{I}$, all the energies of the bulk states are real except for two imaginary-energy edge modes shown in the topologically nontrivial region. Phase $\mathrm{III}$ ($\mathrm{II}$) denotes the $\mathcal{PT}$-symmetry (un)broken regime with the topologically trivial property. Phase $\mathrm{V}$ ($\mathrm{IV}$) denotes the $\mathcal{PT}$-symmetry (un)broken regime with a pair of real-energy edge modes in the topologically nontrivial region. In phase $\mathrm{VI}$, both the bulk and boundary parts are $\mathcal{PT}$-symmetry broken.

\end{document}